\documentstyle[12pt]{article}

\newlength{\dinwidth}
\newlength{\dinmargin}
\newfont\goth{eufm10 scaled 1200}
\newfont\bbl{msbm10 scaled 1095}
\newfont\bbs{msbm10 scaled 1000}
\newfont\bbss{msbm9 scaled 1000}
\newfont\bbsss{msbm7 scaled 1000}
\def\C{\mbox{\bbl C}}

\def\Z{\mbox{\bbl Z}}

\def\ad{{\rm ad}\,}
\def\d{\delta}
\def\1{{\bf 1}}
    \def\f{\varphi}

\renewcommand{\L}{\mbox{$\lambda$}}
\def\Lh{\mbox{$\Lambda$}}
\def\F{\mbox{${\cal F}_\lambda$}}

\def\v{\mbox{$\omega$}}

\def\w{\mbox{$\f$}}
\def\u{\mbox{$f_k\w$}}
\def\P{\mbox{$\Psi$}}
\def\Q{\mbox{$\Phi$}}
\def\x{\otimes}

\def\g{\mbox{\goth g}}

\def\ah{\mbox{$\hat{A}$}}
\def\gah{\mbox{$\hat{\g}(\hat{A})$}}
\def\ga{\mbox{$\g(A)$}}
\def\n{\mbox{\goth n}}
\def\r{\mbox{\goth r}}

\def\h{\mbox{\goth h}}
\def\U{\mbox{\goth U}}
\def\TF{\mbox{{\goth T}(\F)}}
\def\cah{\mbox{$\hat{\mbox{\goth h}}(\hat{A})$}}

\def\su{\mbox{\goth su}}
\newtheorem{satz}{Proposition}
\newtheorem{theo}{Theorem}
\newtheorem{defi}{Definition}

\def\beq{\begin{equation}}      \def\eeq{\end{equation}}
\def\baro{\begin{eqnarray*}}    \def\barr{\begin{eqnarray}}
\def\earo{\end{eqnarray*}}      \def\earr{\end{eqnarray}}
\newcommand{\bensatz}{\begin{satz} {\bf:} \begin{enumerate}}
\newcommand{\bsatz}{\begin{satz} {\bf:}\\ }
\newcommand{\esatz}{\end{satz}}
\newcommand{\beenden}{\end{enumerate}}
\newcommand{\beginnen}{\begin{enumerate}}

 \newcommand{\lb}[1]{\label{#1}}

\newcommand{\If}[1]{\quad\mbox{if #1}}
\newcommand{\by}[1]{\quad\mbox{by #1}}
\newcommand{\for}[1]{\quad\mbox{for #1}}

\begin{document}
\thispagestyle{empty}
\renewcommand{\thefootnote}{\fnsymbol{footnote}}
\begin{flushright} hep-th/9310139 \end{flushright}
\vspace*{2cm}
\begin{center}
{\LARGE \sc On the Fundamental Representation of \\
            Borcherds Algebras with One Imaginary \\[3.5mm]
            Simple Root }\\   
 \vspace*{1cm}
       {\sl Reinhold W. Gebert\footnote[2]{Supported by
        Konrad-Adenauer-Stiftung e.V.}}\\
       {\sl J\"org Teschner\footnote[3]{Supportet by Deutsche
        Forschungsgemeinschaft}}\\
 \vspace*{6mm}
     IInd Institute for Theoretical Physics, University of Hamburg\\
     Luruper Chaussee 149, D-22761 Hamburg, Germany\\
 \vspace*{6mm}
     October 21, 1993\\
\vspace*{1cm}
\begin{minipage}{11cm}\footnotesize
Borcherds algebras represent a new class of Lie algebras which have
almost all the properties that ordinary Kac-Moody algebras have, and
the only major difference is that these generalized Kac-Moody
algebras are allowed to have imaginary simple roots. The simplest
nontrivial examples one can think of are those where one adds ``by
hand'' one imaginary simple root to an ordinary Kac-Moody algebra. We
study the fundamental representation of this class of examples and
prove that an irreducible module is given by the full tensor algebra
over some integrable highest weight module of the underlying Kac-Moody
algebra. We also comment on possible realizations of these
Lie algebras in physics as symmetry algebras in quantum field theory.
\end{minipage}
\end{center}
\renewcommand{\thefootnote}{\arabic{footnote}}
\setcounter{footnote}{0}
\vspace*{1cm}
\section{Introduction}
Studying the Lie algebra of physical states for the 26-dimensional
bosonic string compactified on a torus, Borcherds discovered his
celebrated fake Monster Lie algebra as the first generic example
of a generalized Kac-Moody algebra (cf. \cite{Borc86},
\cite{Borc92} or the review \cite{Gebe93} for physicists). Up to this
point it was only known that the tachyonic ground states give rise to
an infinite rank Lie algebra $L_\infty$ with a set of simple roots
isometric to the Leech lattice and with certain bounds on the dimension
of the root spaces coming from the ``no-ghost'' theorem (cf.
\cite{Con83}, \cite{BoCoQuSl84}, \cite{Fren85}, \cite{GodOli85}). It
was Borcherds' great achievement to observe that this upper bound can
be satisfied by adding a certain set of photonic physical states as
additional generators to the set of generators for $L_\infty$
\cite{Borc90}. Mathematically speaking, he adjoined a set of {\it
imaginary simple roots} (where `imaginary' means `negative norm') to
the set of real simple roots for the Kac-Moody algebra $L_\infty$. In
the sequel Borcherds was able to axiomatize his ideas and he
developed a theory of generalized Kac-Moody algebras in terms of
generators and relations (cf. \cite{Borc88}, \cite{Borc91}).

To get a grasp of these new Lie algebras Slansky \cite{Slan91}
investigated the Borcherds extensions of the Lie algebras $\su(2)$,
$\su(3)$, and affine $\su(2)$ by a single lightlike ($\equiv$ norm zero)
simple root. Computer calculations of the first few weight
multiplicities of the basic representations suggested that the latter
might be written as the tensor algebra over some module for the
underlying nonextended Kac-Moody algebras. In the following we shall
prove that this is true for any Kac-Moody algebra extended by an
arbitrary imaginary simple root.
\section{Definitions}
Let us begin with a review of the definition of Borcherds algebras.
As already mentioned, the original references on the subject are
\cite{Borc92}, \cite{Borc88} and \cite{Borc91}.
\begin{defi} \ \lb{d1} \\
Let $\ah=(a_{ij})$ be a real symmetric $n\times n$ matrix
satisfying the following properties:
\beginnen \item[(i)] either $a_{ii}=2$ or $a_{ii}\le0$,
          \item[(ii)] $a_{ij}\le0$ if $i\neq j$,
          \item[(iii)] $a_{ij}\in\Z$ if $a_{ii}=2$. \beenden
Then the {\bf Borcherds algebra (generalized Kac-Moody algebra)}
associated to \ah\ is defined to be the Lie algebra \gah\ given by the
following generators and relations:\\
Generators: Elements $e_i,\ f_i,\ h_i$ for every $i$; \\
Relations:
\beginnen \item[(0)] $[h_i,h_j]=0$,
          \item[(1)] $[e_i,f_j]=\d_{ij}h_i$,
          \item[(2)] $[h_i,e_j]=a_{ij}e_j,\
                      [h_i,f_j]=-a_{ij}f_j$,
          \item[(3)] $e_{ij}:=(\ad e_i)^{1-a_{ij}}e_j=0,\quad
                      f_{ij}:=(\ad f_i)^{1-a_{ij}}f_j=0
                      \If{}a_{ii}=2\mbox{ and }i\neq j$,
\item[(4)] $e_{ij}:=[e_i,e_j]=0,\quad
                     f_{ij}:=[f_i,f_j]=0\If{}a_{ii}\le0,a_{jj}\le0
                 \mbox{ and }a_{ij}=0$. \beenden
\end{defi}
The elements $h_i$ form a basis for an abelian subalgebra of \gah,
called its {\bf Cartan subalgebra} \cah.
\gah\ has a triangular decomposition
\[ \gah=\hat{\n}_-\oplus\hat{\h}\oplus\hat{\n}_+, \]
where $\hat{\n}_-$ (resp. $\hat{\n}_+$) is the algebra obtained by
dividing the free algebra $\tilde{\n}_-$ ($\tilde{\n}_+$)
generated by the $f_i$ ($e_i$) by the
ideal $\r_-$ ($\r_+$) generated by the $f_{ij}$ ($e_{ij}$).

Note that if $a_{ii}=2$ for
all $i$ then \gah\ is the same as the ordinary Kac-Moody algebra with
symmetrized Cartan matrix \ah. In general, \gah\ has almost all the
properties that ordinary Kac-Moody algebras have, and the only major
difference is that generalized Kac-Moody algebras are allowed to have
imaginary simple roots. In what follows we will exclusively deal with
the case of a Borcherds algebra with one imaginary simple root.

It is clear that if we delete in \ah\ the row and the column
corresponding to the imaginary root then the resulting submatrix $A$
is a generalized Cartan matrix in the sense of Kac \cite{Kac90} with
associated Kac-Moody algebra \ga. Recall the triangular decomposition
\[ \ga=\n_-\oplus\h\oplus\n_+ \]
and the induced decomposition of the universal enveloping algebra:
\[ \U(\ga)=\U(\n_-)\x\U(\h)\x\U(\n_+). \]
An irreducible \ga-module \F\ is called {\bf integrable
  highest-weight module} if there exists a dominant integral weight
$\L\in\h^*$ and a nonzero vector $\v\in\F$ such that
\baro   h(\v)&=&\L(h)\v\for{}h\in\h, \\
     \n_+(\v)&=&0,\\
   \U(\n_-)(\v)&=&\F.        \earo
We denote by \TF\ the tensor algebra over \F,
\[ \TF:=\bigoplus_{n=0}^\infty\F^n\equiv
   \C\cdot\1\oplus\F\oplus(\F\x\F)\oplus(\F\x\F\x\F)\oplus\ldots \]
Now we are ready to state our result.
\section{The theorem}
\begin{theo} \ \lb{t1} \\
Let $\ah=(a_{ij})$, $0\le i,j\le n$, be a symmetric integer matrix
satisfying the following properties:
\beginnen \item[(i)] $a_{00}\le0$, $a_{ii}=2$ for $1\le i\le n$,
          \item[(ii)] $a_{ij}\le0$ if $i\neq j$.
\beenden
Let \F\ be the integrable highest weight module over the Kac-Moody
algebra \ga\ associated to the Cartan matrix $A=(a_{ij})$,
$1\le i,j\le n$, with highest weight \L\ defined by
$\L(h_i):=-a_{0i}$, $1\le i\le n$, and highest weight vector \v.
Then the tensor algebra \TF\ over \F\ is
\gah-module isomorphic to the highest weight module $L(\Lambda)$,
$\Lh(h_i)=\d_{i0}$, $0\le i\le n$ of \gah .
\end{theo}
Proof:\\
We define an action of the generators of \gah\ on the tensor algebra
\TF\ as follows. Our convention for indices will be that $i,j,k$ run
from 1 to $n$\ unless otherwise stated!\\
      The Kac-Moody generators $e_i,h_i,f_i$ act trivially on the
      ``vacuum'' vector \1 and as highest weight representation
      on \F. We extend this action to the tensor algebra \TF\ by
      Leibnitz' rule. \\
      The generator $h_0$ acts diagonal:
      \barr h_0(\1)&:=&\1, \lb{h1} \\
            h_0(\v)&:=&(1-a_{00})\v, \lb{hv} \\
            h_0(\u)&:=&-a_{0k}\u+f_kh_0(\w)\for{}\w\in\F, \lb{hu} \\
            h_0(\Q\x\P)&:=&h_0(\Q)\x\P+\Q\x h_0(\P)-\Q\x\P
                           \for{}\Q,\P\in\TF. \lb{hx} \earr
      The ``imaginary'' generator $f_0$ adjoins one tensor factor of
      the highest weight vector \v, i.e.,
      \beq  f_0(\P):=\v\x\P\for{}\P\in\TF. \lb{fx} \eeq
      For $e_0$ we put
      \beq e_0(\1):=0, \lb{e1} \eeq
      while for the definition on $\F^n$, $n\ge1$, we observe that
      $\F^n= \U(\n_-)(\v\x\F^{n-1})$, so that it is sufficient to
      require, inductively,
      \barr e_0(f_i(\P))&:=&f_i(e_0(\P)), \lb{ef} \\
            e_0(\v\x\P)&:=&h_0(\P)+\v\x e_0(\P), \lb{ex} \earr
      for $\P\in\F^n$, $n\ge0$.

Having defined the action of the generators on the tensor algebra
we will now check that \TF\ carries the claimed \gah-module
structure. First we note that $h_0$ and the $h_i$'s are defined to
act diagonal on the tensor algebra. Hence the $h$'s commute with each
other. Secondly, all commutation relations involving only Kac-Moody
generators $e_i,h_i,f_i$ are valid by assumption. Next, we
have a look at those commutation relations which are more or less
trivial since they can be checked immediately on the whole tensor
algebra,
\baro (e_0f_0-f_0e_0)(\P)&=&e_0(\v\x\P)-\v\x e_0(\P)=h_0(\P), \\
      (e_0f_i-f_ie_0)(\P)&=&0, \\
      (e_if_0-f_0e_i)(\P)&=&e_i(\v\x\P)-\v\x e_i(\P) =0, \\
      (h_0f_0-f_0h_0)(\P)&=&h_0(\v\x\P)-\v\x h_0(\P) \\
                         &=&(h_0-1)(\v)\x\P \\
                         &=&-a_{00}f_0(\P), \\
      (h_if_0-f_0h_i)(\P)&=&h_i(\v\x\P)-\v\x h_i(\P) \\
                         &=&h_i(\v)\x\P \\
                         &=&-a_{0i}f_0(\P). \earo
Finally we check the remaining four types of commutators:
\baro (h_0f_i-f_ih_0)(\1)&=&-f_i(\1)=0=-a_{0i}f_i(\1), \\
      (h_0f_i-f_ih_0)(\w)&=&-a_{0i}f_i\w, \\
      (h_0f_i-f_ih_0)(\Q\x\P)&=&h_0(f_i(\Q)\x\P+\Q\x f_i(\P)) \\
                       &\ &\quad-f_i(h_0(\Q)\x\P+\Q\x h_0(\P)-\Q\x\P) \\
                         &=&(h_0f_i-f_ih_0)(\Q)\x\P
                                             +\Q\x(h_0f_i-f_ih_0)(\P) \\
                         &=&-a_{0i}f_i(\Q\x\P) \by{induction}, \\[1cm]
      (h_0e_i-e_ih_0)(\1)&=&0=a_{0i}e_i(\1), \\
      (h_0e_i-e_ih_0)(\v)&=&0=a_{0i}e_i(\v), \\
      (h_0e_i-e_ih_0)(\u)&=&h_0(\d_{ik}h_i(\w)+f_ke_i(\w))
                           -e_i(-a_{0k}\u+f_kh_0(\w)) \\
                          &=&a_{0k}(e_if_k-f_ke_i)(\w)
                           +f_k(h_0e_i-e_ih_0)(\w) \\
                          &=&a_{0k}\d_{ik}h_i(\w)+a_{0i}f_ke_i(\w)
                             \by{induction} \\
                          &=&a_{0i}e_i(\u), \\
      (h_0e_i-e_ih_0)(\Q\x\P)&=&h_0(e_i(\Q)\x\P+\Q\x e_i(\P)) \\
                     &\ &\quad-e_i(h_0(\Q)\x\P+\Q\x h_0(\P)-\Q\x\P) \\
                          &=&(h_0e_i-e_ih_0)(\Q)\x\P
                           +\Q\x(h_0e_i-e_ih_0)(\P) \\
                          &=&a_{0i}e_i(\Q\x\P) \by{induction}, \\[1cm]
      (h_ie_0-e_0h_i)(\1)&=&0=a_{i0}e_0(\1), \\
      (h_ie_0-e_0h_i)(\v)&=&h_i(\1)+a_{0i}e_0(\v)=a_{0i}\1
                          =a_{i0}e_0(\v), \\
      (h_ie_0-e_0h_i)(\u)&=&0=a_{i0}e_0(\u), \\
      (h_ie_0-e_0h_i)(\v\x\P)&=&h_i(h_0(\P)+\v\x e_0(\P))
                           -e_0(h_i(\v)\x\P+\v\x h_i(\P)) \\
                          &=&a_{0i}h_0(\P)+\v\x(h_ie_0-e_0h_i)(\P)
                             +(h_ih_0-h_0h_i)(\P) \\
                          &=&a_{i0}e_0(\v\x\P), \\
      (h_ie_0-e_0h_i)(\u\x\P)&=&h_i(-e_0(\w\x f_k(\P))
                                                   +f_k(e_0(\w\x\P))) \\
                       &\ &\quad-e_0(h_i(\u)\x\P+\u\x h_i(\P)) \\
                     &=&-h_i(e_0(\w\x f_k(\P)))+h_i(f_k(e_0(\w\x\P))) \\
             &\ &\quad-a_{ik}e_0(\w\x f_k(\P))+a_{ik}f_k(e_0(\w\x\P)) \\
              &\ &\quad+e_0(h_i(\w)\x f_k(\P))-f_k(e_0(h_i(\w)\x\P)) \\
              &\ &\quad+e_0(\w\x f_k(h_i(\P)))-f_k(e_0(\w\x h_i(\P))) \\
                          &=&(e_0h_i-h_ie_0)(\w\x f_k(\P)) \\
              &\ &\quad+f_k((h_ie_0-e_0h_i)(\w\x\P)) \\
                          &=&a_{i0}(-e_0(\w\x f_k(\P))
                             +f_k(e_0(\w\x\P))) \by{induction} \\
                          &=&a_{i0}e_0(\u\x\P), \\[1cm]
      (h_0e_0-e_0h_0)(\1)&=&0=a_{00}e_0(\1), \\
      (h_0e_0-e_0h_0)(\v)&=&h_0(\1)-(1-a_{00})e_0(\v)
                          =a_{00}\1=a_{00}e_0(\v), \\
      (h_0e_0-e_0h_0)(\u)&=&0=a_{00}e_0(\u), \\
      (h_0e_0-e_0h_0)(\v\x\P)&=&h_0(h_0(\P)+\v\x e_0(\P))
                                -e_0(-a_{00}\v\x\P+\v\x h_0(\P)) \\
                          &=&a_{00}h_0(\P)
                                +\v\x(h_0e_0-e_0h_0)(\P) \\
                          &=&a_{00}e_0(\v\x\P) \by{induction}, \\
      (h_0e_0-e_0h_0)(\u\x\P)&=&h_0(-e_0(\w\x f_k(\P))
                                                   +f_k(e_0(\w\x\P))) \\
                       &\ &\quad-e_0(h_0(\u)\x\P+\u\x h_0(\P)-\u\x\P) \\
                     &=&-h_0(e_0(\w\x f_k(\P)))+h_0(f_k(e_0(\w\x\P))) \\
             &\ &\quad-a_{0k}e_0(\w\x f_k(\P))+a_{0k}f_k(e_0(\w\x\P)) \\
              &\ &\quad+e_0(h_0(\w)\x f_k(\P))-f_k(e_0(h_0(\w)\x\P)) \\
              &\ &\quad+e_0(\w\x f_k(h_0(\P)))-f_k(e_0(\w\x h_0(\P))) \\
              &\ &\quad+e_0(\w\x f_k(\P))-f_k(e_0(\w\x\P)) \\
                          &=&(e_0h_0-h_0e_0)(\w\x f_k(\P)) \\
              &\ &\quad+f_k((h_0e_0-e_0h_0)(\w\x\P)) \\
                             &=&a_{00}(-e_0(\w\x f_k(\P))
                                +f_k(e_0(\w\x\P))) \by{induction} \\
                             &=&a_{00}e_0(\u\x\P), \earo
for all $\w\in\F$ and $\Q,\P\in\TF$.

Now we shall prove that \TF\ is indeed
isomorphic to $L(\Lambda)$ as a \gah -module. Denote the highest
weight vector of $L(\Lambda)$ by $v_{\Lambda}$.
Define a map $\nu:\U(\tilde{\n}_-)v_{\Lambda}\rightarrow\TF$ by
\[ \nu(f_{i_1}\ldots f_{i_n}v_{\Lambda})
:=f_{i_1}\ldots f_{i_n}(\1) \]
where $i_1\ldots i_n \in \{0,\ldots ,n\}$, and linearity.
To prove that $\nu$ reduces to a well defined \gah-module homomorphism
$\nu':\U(\hat{\n}_-)v_{\Lambda}\rightarrow\TF$,
one has to check that the action of
elements of $\r_-$ on \TF\ vanishes, i.e. that
the Serre-relations are valid.
For $f_{ij},i,j=1\ldots n$, this is part of the definition. To check
the remaining ones, observe that
\[ ((\ad f_i)^m f_0)(\P)=f_i^m\v\otimes\P, \]
so that for $i=1\ldots n$
\[ f_{i0}(\P)=f_i^{1+\lambda (h_i)}\v\otimes\P=0 \]
because of lemma 10.1 of \cite{Kac90}. According to \cite{Borc88}
(see als \cite{Har})
the irreducible module $L(\Lambda)$ is obtained from the Verma-module
$M(\Lambda)$ by dividing out the subspace generated by the primitive
vectors $f_i^{1+\Lambda (h_i)}v_{\Lambda}, i=1\ldots n$.
Because of $f_i^{1+\Lambda (h_i)}(\1)=f_i(\1)=0$,
$\nu'$ reduces further
to a map $\nu'':L(\L)\rightarrow \TF$.
$\nu''$ is injective because the kernel of $\nu''$ would be a proper
submodule of $L(\L)$, and surjective because \TF\ is spanned by
vectors of the form
\[ u_1\v\otimes\ldots\otimes u_n\v=\nu(u_1(f_0)\ldots u_n(f_0)
v_{\Lambda}) \]
where \begin{eqnarray*}
u_i=F_{n_1(i)}\ldots F_{n_{k(i)}(i)}, & F_{n_j(i)} \in \ga, &
u_i(f_0)=\left[F_{n_1(i)},\left[\ldots\left[
F_{n_{k(i)}(i)},f_0\right]\ldots\right]\right]. \end{eqnarray*}

We observe that the theorem is not altered if we replace $a_{00}$
by any nonpositive real number or $\Lh(h_0)$ by any positive real
number.
\section{Outlook}
According to a conjecture of Ginsparg \cite{Slan91}, the special class
of Borcherds algebras considered in the theorem might play a role in
second quantization of a single particle theory. In this interpretation
we regard the module \F\ from above as one-particle Fock space so
that \TF\ comprises all multiparticle states. In other words, within
a single irreducible represenation of the Borcherds algebra we
encounter all possible multiparticle excitations. Thus the
``imaginary'' generators $f_0$ and $e_0$ act as particle creation and
particle annihilation operators, respectively, whereas the vector \1\
indeed deserves the name ``true vacuum'' in contrast to the ``ground
state'' $\v\in\F$.

Applying this idea to string theory one should think about the
underlying Kac-Moody algebra \ga\ as spectrum generating algebra for
the physical states of the bosonic string. Consequently, the tensor
algebra \TF\ would be intimately related to a string field theory.
Note that in the special case of an underlying affine Lie algebra \ga\,
we would end up with a string field theory on the group manifold
associated to \ga\ (cf. \cite{GepWit86}).

It is clear that the emergence of Borcherds algebras in quantum field
theory is just a naive speculation since up to now at least one
important point in dealing with particles is missing. The tensor
algebra \TF\ carries no symmetry or antisymmetry constraints at all,
which means that the concept of statistics is absent. At present,
work is in progress to clarify how symmetrization of \TF\ can be
implemented into \gah\ algebraically via additional relations.

In view of these possible realizations of Borcherds algebras in physics
we shall finish with the useful construction of a ``number operator''
which counts the number of $f_0$'s (number of particles/strings)
occurring in the expression for a homogeneous state vector $\P\in\TF$.
We are looking for an element $N$ in the Cartan subalgebra \cah\
satisfying
\[ N(\P)\stackrel!=n\P\qquad\forall\P\in\F^n,n\ge1, \]
or, equivalently,
\[ [N,f_j]\stackrel!=\d_{j0}f_j\for{}0\le j\le n. \]
The ansatz $N=\sum_{i=0}^nN_ih_i$ yields the following system of linear
equations for the rational coefficients $N_i$:
\[ \sum_{i=0}^na_{ij}N_i\stackrel!=-\d_{j0}\for{}0\le j\le n. \]
If \ah\ is invertible we obtain a unique solution for the number
operator $N$. Note, however, that the eigenvalues of $N$ give us the
number of $f_0$'s shifted by $N_0$ since we have $N(\1)=N_0\1$
instead of $N(\1)=0$. This annoying constant may be removed by
defining the ``renormalized'' number operator $\hat{N}:= N-N_0$.

\end{document}